\documentclass[runningheads]{llncs}
\usepackage{graphicx}
\usepackage{caption}
\usepackage{subcaption}
\usepackage{xcolor}
\usepackage{cite}
\usepackage[normalem]{ulem}

\begin{document}

\title{Piloting topic-aware research impact assessment features in BIP! Services}

%
%

\author{Serafeim Chatzopoulos\inst{1}\orcidID{0000-0003-1714-5225} \and
Kleanthis Vichos\inst{1}\orcidID{0000-0002-8955-9489} \and
Ilias Kanellos\inst{1}\orcidID{0000-0003-2146-3795} \and Thanasis Vergoulis\inst{1}\orcidID{0000-0003-0555-4128}}
\authorrunning{Chatzopoulos, Vichos, et al.}

%
\institute{IMSI, Athena RC, Athens, Greece\\
\email{\{schatz, kvichos, ilias.kanellos, vergoulis\}@athenarc.gr}}
\maketitle              
\begin{abstract}
Various research activities rely on citation-based impact indicators. However these indicators are usually globally computed, hindering their proper interpretation in applications like research assessment and knowledge discovery. In this work, we advocate for the use of topic-aware categorical impact indicators, to alleviate the aforementioned problem. In addition, we extend BIP! Services to support those indicators and showcase their benefits in real-world research activities.
\end{abstract}
\section{Introduction}
\label{sec:intro}

Citation-based impact indicators, like citation counts, have found a variety of uses during the previous years as a way to facilitate various research-related activities. First of all, they 
are used by scientific literature search engines (e.g., Semantic Scholar\footnote{Semantic Scholar: \url{https://www.semanticscholar.org/}}, BIP Finder\footnote{BIP! Finder: \url{https://bip.imsi.athenarc.gr/} }) to rank keyword search results assisting researchers in prioritising their reading. 
Moreover, they have been exploited as facilitators in research assessment activities~\cite{ten-ways}, while they have also become the basis for 
monitoring scientific output (e.g.,~\cite{papastefanatos2020open}). 
The majority of these indicators are based on network analysis algorithms that 
rely on citation data and publication metadata (e.g., publication year, author lists etc). 

However, impact indicators have been related to various problems that plague research community at large. For instance, scientific literature search engines incorporate a limited number of indicators that capture a narrow perspective of scientific impact~\cite{vergoulis2019bip}. 
Specifically, most of them only support citation count, which has specific known issues (e.g., bias against recent articles, vulnerable to excessive self citation attacks). 
Moreover, in research assessment, evaluators often tend to over-rely on impact indicators without delving into the researchers' CVs and publications. 
Using indicators as ``evaluation shortcuts'' has been identified as a problematic approach~\cite{ten-ways} that often results in unfair
research assessment. 

But an even more important problem is 
that, in the aforementioned applications, 
users are allowed to compare articles from different fields, something that can lead to misconceptions. 
Academic search engines often return results from different topics, since the same keywords can be related to various fields. Similarly, academic CVs usually contain publications from multiple fields, hence directly comparing impact indicators likely results in misjudgements.

Since it is not realistically possible to alleviate all impact-indicator-related problems, 
they should always be used with caution and only supplementary to other (qualitative) evidence. However, impact indicators can still have an assisting role in various applications, 
therefore alleviating some of their problems remains valuable. 
Motivated by this, we adapt the multi-perspective impact indicators provided by BIP! DB~\cite{vergoulis2021bip} into a set of topic-aware, categorical indicators. To transform the numerical values of the original indicators into categorical, we translate them into percentile rank classes (similarly to the approaches described in \cite{bornmann2013use}). We believe that this is useful since the categorical indicators are easier to interpret. Finally, we showcase the benefits of these topic-aware indicators in real-world applications by extending the BIP! Services\footnote{BIP! Services: \url{https://bip.imsi.athenarc.gr/}} to incorporate them.

\section{Implementation}
\label{sec:approach}

\subsection{Topic-aware, categorical impact indicators}

As mentioned, to alleviate the problems mentioned in Section~\ref{sec:intro}, we advocate on the use of a variation of the impact indicators offered by BIP! DB~\cite{vergoulis2021bip}. This database already follows a multi-perspective approach providing a variety of indicators that capture different aspects of publication scientific impact. For this work we focus on the following indicators:\footnote{More details (e.g., the calculation algorithms) for these indicators can be found here: \url{https://bip.imsi.athenarc.gr/site/indicators}}
\begin{itemize}
    \item \emph{Popularity.} It reflects the ``current" impact/attention (the ``hype") of an article based on the underlying citation network.
    \item \emph{Influence.} It reflects the overall/diachronic impact of an article in the research community at large, based on the underlying citation network.
    \item \emph{Citation Count.} The number of citations an article has received (it also reflects overall/diachronic impact).
    \item \emph{Impulse.} It reflects the initial momentum of an article directly after its publication, based on the underlying citation network.
\end{itemize}

BIP! DB calculates scores on the whole citation network. 
Based on the indicator value, it is possible to assign a global categorical value to each paper according to the percentile\footnote{Percentiles are not strongly affected by outlier values, and can be easily calculated even if the underlying data are heavily skewed.} into which it belongs. 
Hence, in this way, it is possible to define five categorical impact indicators, one for each of the initial indicators. We refer to these categorical indicators as ``Popularity class'', ``Influence class'', ``Citation count class'', and ``Impulse class'', respectively. 

We proceed a step further by annotating each 
article with 
its
relevant topics (details in Section~\ref{sec:data}) and, then, calculating topic-specific versions of the aforementioned categorical indicators. 
In this way, for each article, apart from its global impact classes we also calculate topic-specific ones for all 
its related topics. 
In particular, for each topic, we compute the percentiles for all indicators as follows:
first, we rank the topic-related articles by the given impact indicator in descending order;
then, each publication is assigned a percentile based on the distribution of scores and we assign the respective class to the article.
For all categorical indicators, the following impact classes are used: \texttt{Top 0.01\%}, \texttt{Top 0.1\%}, \texttt{Top 1\%}, \texttt{Top 10\%}, \texttt{Average} (rest $90$\%).

\subsection{Data collection, processing, and publishing}
\label{sec:data}

To calculate the topic-aware, categorical impact indicators, we get the respective impact indicator scores from BIP! DB. The version used for the needs of the current paper was version~$8$ containing indicators for almost $134$M articles.\footnote{\url{https://doi.org/10.5281/zenodo.4386934}} We then associated these articles with (L2) topics from OpenAlex~\cite{openalex} ($284$ in total). 
We chose to keep only the three most dominant topics for each publication, based on their confidence score, and only if this score was greater than $0.3$.
After this process, 
we ended up with more than $75$\% of the articles in BIP! DB being associated to at least one topic. Subsequently, we calculated the topic-specific impact classes for each publication.
Given a specific topic, each publication was assigned with an impact class from the set \texttt{\{C1, C2, C3, C4, C5\}}, with \texttt{C1} corresponding to the \texttt{Top 0.01\%} class and \texttt{C5} to that of \texttt{Average} impact.
We integrated those indicators in BIP! DB dataset that is openly available on Zenodo.

\subsection{BIP! Services extensions}

\begin{figure}[t]
    \centering
    \begin{subfigure}{.47\columnwidth}
        \centering
        \fbox{\includegraphics[width=1\columnwidth]{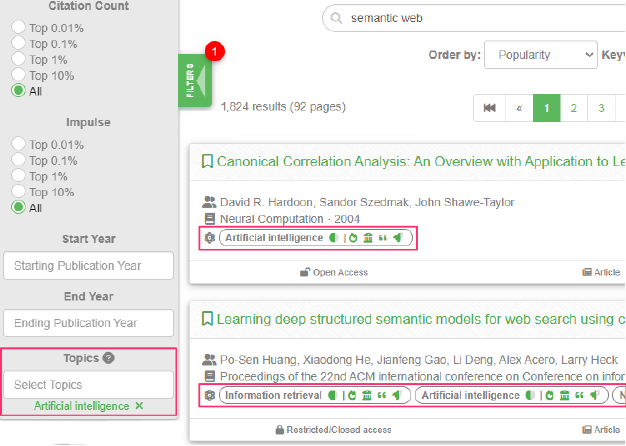}}%
        \caption{Scientific knowledge discovery.}
    \end{subfigure}\hfill%
    \begin{subfigure}{.47\columnwidth}
        \centering
        \fbox{\includegraphics[width=1\columnwidth]{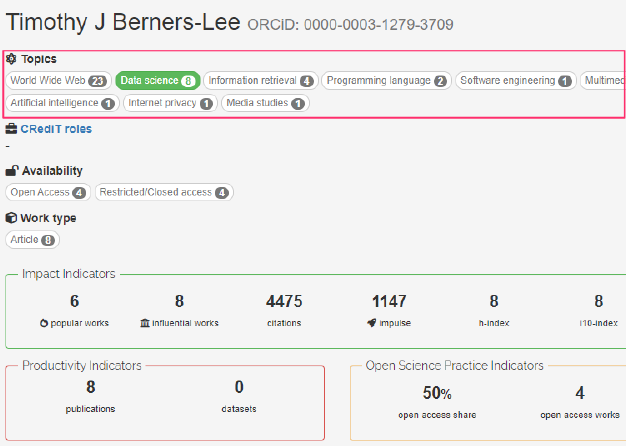}}%
        \caption{Monitoring a researcher's impact.}
    \end{subfigure}\hfill%
    \vspace{-2mm}
    \caption{Topic-aware publication impact indicators in BIP! Services.}
    \vspace{-5mm}
    \label{fig:use-cases}
\end{figure}

To demonstrate how the previous indicators can be useful in practice, we focused on two use-cases:
scientific knowledge discovery and research impact monitoring. For the former case, we have extended the BIP! Finder~\cite{vergoulis2019bip} academic search engine accordingly by modifying the UI to (a) display the topics of each result and its impact class according to each topic and (b) support topic-based filtering. 
To visualise the impact classes, we have used a compact visualisation based on icons that get particular color-codes for each class. Figure~\ref{fig:use-cases}a illustrates the results list and the filter for the query ``semantic web''. 
For the latter case, we have extended the BIP! Scholar~\cite{vergoulis2022bip} service that offers researcher profile pages summarising research careers. Specifically, we have added a topic facet allowing the researchers to reveal their impact on selected topics (Figure~\ref{fig:use-cases}b).

\section{Demonstration scenarios}

At the conference, the audience will have the opportunity to interact with the BIP! Services and examine the benefits that the topic-specific impact indicators bring in various use-cases. We will also demonstrate the following scenarios. 

\noindent \textbf{Scientific knowledge discovery.}
An audience member searches for the keywords ``semantic web" in BIP! Finder and determines (using the topic filter) that only articles related to the ``Artificial Intelligence'' topic are of interest (Figure~\ref{fig:use-cases}a). Each result contains the associated topics and for each of them the impact icons inform the user about the topic-specific impact class of the result. 
 
\noindent \textbf{Monitoring a researcher's impact.} 
The same audience member, uses BIP! Scholar, to display the  profile of Tim Berners-Lee, a well-known researcher in field of web technologies (Figure~\ref{fig:use-cases}b). By selecting each of the topic facets on top of the profile (``Data science''), the user can reveal the impact of Berners-Lee in the respective topic (e.g., how many popular works he has). 

\section*{Acknowledgements}
This work was co-funded by the EU Horizon Europe projects SciLake (GA: 101058573) and GraspOS (GA: 101095129)
and the EU H2020 project OpenAIRE-Nexus (GA: 101017452).

\begin{figure}[!h]
     \centering
         \includegraphics[width=0.1\linewidth]{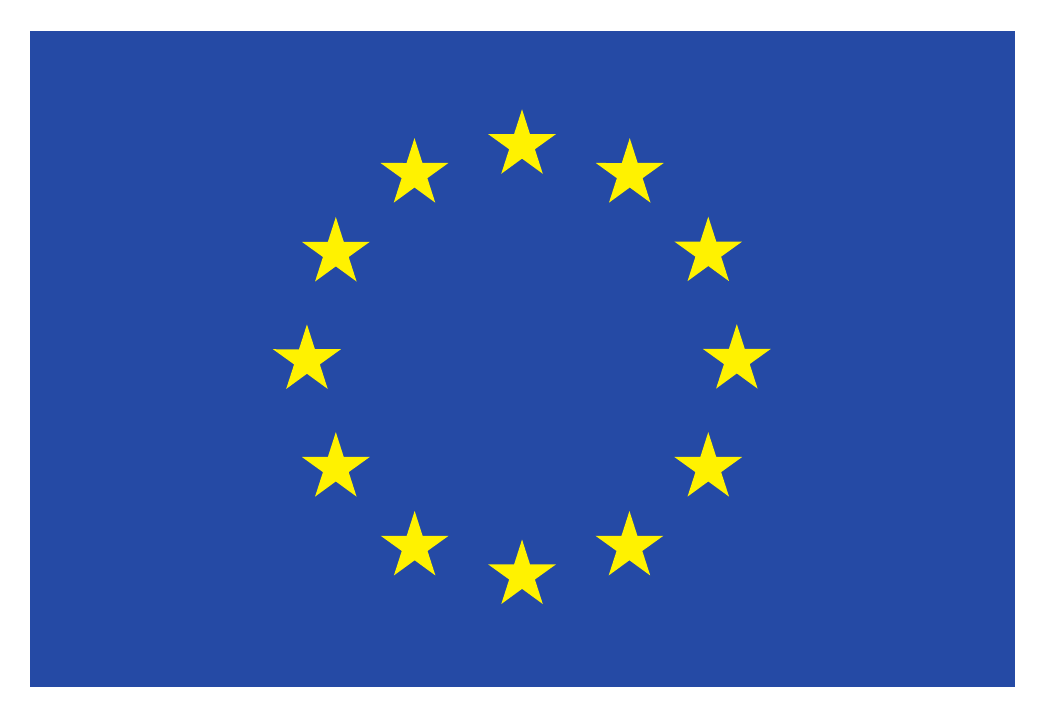}
\end{figure}
\bibliographystyle{splncs04}
\bibliography{refs}

\end{document}